\documentclass[12pt]{iopart}
\usepackage{cite}
\usepackage{graphicx}
\usepackage{iopams}
\usepackage{bm}
\usepackage{amssymb}
\usepackage[usenames,dvipsnames,svgnames,table]{xcolor}
\usepackage[unicode=true,
            pdfusetitle,
            bookmarks=true,
            bookmarksnumbered=false,
            bookmarksopen=false,
            breaklinks=true,
            pdfborder={0 0 0},
            backref=false,
            colorlinks=true]{hyperref}
\hypersetup{linkcolor=NavyBlue,urlcolor=NavyBlue,citecolor=NavyBlue}

\begin{document}

\title{Valid lower bound for all estimators in quantum parameter estimation}

\author{Jing Liu}
\address{Department of Mechanical and Automation Engineering, The Chinese University of Hong Kong, Shatin, Hong Kong, China}

\author{Haidong Yuan}
\ead{hdyuan@mae.cuhk.edu.hk}
\address{Department of Mechanical and Automation Engineering, The Chinese University of Hong Kong, Shatin, Hong Kong, China}

\begin{abstract}
The widely used quantum Cram\'er-Rao bound (QCRB) sets a lower bound for the mean square error of unbiased estimators
in quantum parameter estimation, however, in general QCRB is only tight in the asymptotical limit. With limited number of
measurements biased estimators can have far better performance for which QCRB cannot calibrate. Here we introduce a
valid lower bound for all estimators, either biased or unbiased, which can serve as standard of merit for all quantum parameter estimations.
\end{abstract}


\maketitle

\section{Introduction \label{sec:introduction}}

An important task in quantum metrology is to find out the ultimate achievable precision limit and design schemes to attain it. This turns
out to be a hard task, and one often has to resort to various lower bounds to gauge the performance of heuristic approaches,
such as the quantum Cram\'er-Rao bound~\cite{HELS67,HOLE82,BRAU94,BRAU96}, the quantum Ziv-Zakai bound
~\cite{TsangZZ}, quantum measurement bounds~\cite{Giovannetti2012} and Weiss-Weinstein family of error bounds~\cite{Lu2015}.
Among these bounds  the quantum Cram\'er-Rao bound (QCRB) is the most widely used lower bound for unbiased estimators
~\cite{Fisher,CRAM46,Rao,Pezze2006,Pezze2008,Giovannetti2011, GIOV06,Fujiwara2008,
Escher2011,Escher2012,Tsang2013,Rafal2012,Knysh2014,
Toth2014, Yao2015, Xiao2014, Ozaydin2015,Liu2016a,Liu2016b,Lane1993,Braunstein1992,Pezze2013,Pezze2015}.
However, with limited number of measurements many practical estimators are usually biased.
For example the minimum mean square error (MMSE)
estimator, which is given by the posterior mean $\hat{x}(y)=\int p(x|y)x dx$~\cite{Clarkson}, is in general biased in the finite regime,
here $x$ denotes the parameter and $y$ denotes measurement results, the posterior probability distribution $p(x|y)$ can be obtained 
by the Bayes' rule
$p(x|y)=\frac{p(y|x)p(x)}{\int p(y|x)p(x)dx},$
with $p(x)$ as the prior distribution of $x$ and $p(y|x)=Tr(\rho_xM_y)$ given by the Born's rule. The MMSE estimator provides the 
minimum mean square error
\begin{equation}
\mathrm{MSE}(\hat{x})=\int p(x)\sum_{k=0}^{n}(\hat{x}(y)-x)^2p(y|x)~dx. \label{eq:standard}
\end{equation}
The performance of this estimator, however, cannot be calibrated by quantum Cram\'er-Rao bound in the finite regime as with limited 
number of measurements it is usually biased. This is also the case for many other estimators including the commonly used maximum 
likelihood estimator~\cite{Lane1993,Braunstein1992,Pezze2013,Pezze2015}.

In this article we derive an optimal biased bound (OBB) which sets a valid lower bound for all estimators in quantum parameter estimation, either biased or unbiased. This bound works for arbitrary number of measurements, thus can be used to gauge the performances of all estimators in quantum parameter estimation. And the difference between this bound and the quantum Cram\'er-Rao bound also provides a way to gauge when quantum Cram\'er-Rao bound can be safely used, i.e., it provides a way to gauge the number of measurements needed for entering the asymptotical regime that the quantum Cram\'er-Rao bound works. The classical optimal biased bound has been used in classical signal processing~\cite{Young,Eldar2009}.

\section{Main Result}
Based on different assumptions there exists different ways of deriving lower bounds, for example some Bayesian quantum Cram\'er-Rao bound, which based on quantum type of
Van Tree inequality, has been obtained~\cite{Gill95,Yuen1973,Gill2000}. These bounds require the differentiability of the prior distribution at the boundary of the support region, thus may not apply, for example, to the uniform prior distribution. The optimal biased bound does not require the differentiability of the prior distribution at the boundary, thus can be applied more broadly. For the completeness, we will first follow the treatment of Helstrom~\cite{HELS67} to derive a lower bound for estimators with a fixed bias, from which we then derive a valid
lower bound for all estimators by optimizing the bias.

We consider the general case of estimating a function $f(x)$ for the interested parameter $x$ with a given prior distribution. To make
any estimation, one needs to first perform some measurements on the state $\rho_x$, which are generally described by
a set of Positive Operator Valued Measurements (POVM), denoted as $\{\Pi_y\}$. The measurements have probabilistic outcomes $y$ with probability
 $p(y|x)=\mathrm{Tr}(\Pi_y\rho_x)$.
An estimator $\hat{f}(y)$, based on the measurement results $y$, has a mean
$E(\hat{f}(y)|x)=\int\!\hat{f}(y)\mathrm{Tr}(\rho_x\Pi_{y})dy=f(x)+b(x),$
where $b(x)$ represents the bias of the estimation. This equation can be written in another form
\begin{equation}
\int\!\!\left(\hat{f}(y)-E(x) \right)\mathrm{Tr}\left(\rho_x\Pi_y\right)dy=0. \label{eq:unbia}
\end{equation}
where we use $E(x)$ as a short notation for $E(\hat{f}(y)|x)$ which equals to $f(x)+b(x)$ and only depends on $x$.
Assuming the prior distribution is given by $p(x)$, the mean square error is then in the form
\begin{equation}
\mathrm{MSE}(\hat{f}) \!=\!\! \int\!\!dx\!\!\int\!\!p(x)\!\!\left[\hat{f}(y)-f(x)\right]^2 \!\mathrm{Tr}\left(\rho_x\Pi_y\right)dy
\!=\!\! \int\!\!p(x)[\delta \hat{f}^2+b^2(x)]dx,
\label{eq:msedef}
\end{equation}
where $\delta \hat{f}^2=\int (\hat{f}(y)-E(x))^2\mathrm{Tr}(\rho_x\Pi_y)dy$ is the variance of $\hat{f}(y)$.

Differentiating Eq.~(\ref{eq:unbia}) with respect to $x$ and use the fact
that $\int E'(x)\mathrm{Tr}(\rho_x\Pi_y)dy=E'(x)$, with $E'(x):=\partial E(x)/\partial x$, we get
\begin{equation}
\int \!\left(\hat{f}(y)-E(x)\right)\mathrm{Tr}\left(\frac{\partial\rho_x}{\partial x}\Pi_y \right)dy=E'(x). \label{eq:d1}
\end{equation}
Now multiply $p(x)$ at both sides of Eq.~(\ref{eq:d1}) and substitute the following
equation into it
\begin{equation}
\frac{\partial \rho_x}{\partial x}=\frac{1}{2}(\rho_x L+L\rho_x), \label{eq:SLD}
\end{equation}
here $L$ is known as the symmetric logarithmic derivative of $\rho_x$ which
is the solution to Eq.~(\ref{eq:SLD}). We then obtain
\begin{equation}
\mathrm{Re}\!\!\int\!\!p(x)\left(\hat{f}(y)-E(x)\right)\mathrm{Tr}\left(\rho_x L\Pi_y\right)dy=E'(x) p(x), \label{eq:prio1}
\end{equation}
where $\mathrm{Re}(\cdot)$ represents the real part. Multiply both sides again with a real
function $z(x)$ then integrate with respect to $x$,
\begin{equation}
\mathrm{Re}\!\!\int\!dx\!\!\int\!\!p(x)\!\left(\hat{f}(y)-E(x)\right)\!\mathrm{Tr}\left(z(x)\rho_x L\Pi_y\right)dy
=\!\int\!\!p(x)E'(x) z(x)dx.  \label{eq:prior}
\end{equation}
Now we denote $A=\sqrt{p(x)}(\hat{f}(y)-E(x))\sqrt{\rho_x\Pi_y}$ and
$B=\sqrt{p(x)}z(x)\sqrt{\rho_x}L\sqrt{\Pi_y}$,  then the left side of above
equation can be rewritten as $\mathrm{Re}\int\!dx\!\int \mathrm{Tr}(A^{\dagger}B)dy$.
Therefore, Eq.~(\ref{eq:prior}) now has the form
\begin{equation}
\mathrm{Re}\int\!dx\!\int \mathrm{Tr}(A^{\dagger}B)~dy =\int p(x)E'(x) z(x)~dx.  \label{eq:prior2}
\end{equation}
Using Schwarz inequality we have
\begin{eqnarray}
\Big|\mathrm{Re}\int\!\!dx\!\!\int\!\!\mathrm{Tr}(A^{\dagger}B)dy\Big|^2
&\leq& \left(\int\!dx\!\int\!\!\mathrm{Tr}(A^\dagger A)dy\right)\!\!
\left(\int\!dx\!\int\!\! \mathrm{Tr}(B^\dagger B)dy\right) \nonumber \\
&=&\int p(x)\delta \hat{f}^2~dx \int p(x)z^2(x) J(\rho_x)~dx, \nonumber
\end{eqnarray}
the last equality we used the fact that
\begin{eqnarray}
\Big|\int\!dx\!\int \mathrm{Tr}(A^\dagger A) dy\Big|
& &=\Big|\int\!dx\!\int p(x)\left(\hat{f}(y)-E(x)\right)^2\mathrm{Tr}\left(\rho_x\Pi_y\right)dy \Big| \nonumber \\
& &=\int p(x)\delta \hat{f}^2~dx,
\end{eqnarray}
and
\begin{eqnarray}
\Big|\int\!dx\!\int \mathrm{Tr}(B^\dagger B)~dy\Big|
&=\int p(x)z^2(x) \mathrm{Tr}(\rho_x L^2)~dx \nonumber \\
&= \int p(x)z^2(x)J(\rho_x)~dx.
\end{eqnarray}
Here $J(\rho_x)=\mathrm{Tr}(\rho_xL^2)$ is the quantum Fisher information~\cite{HELS67,HOLE82}.
Based on above equations, we can obtain
\begin{equation}
\int p(x)\delta \hat{f}^2~dx\geq \frac{|\int p(x)E'(x) z(x)~dx|^2}{\int p(x)z^2(x)J(\rho_x)~dx},
\end{equation}
which is valid for any $z(x)$ that satisfies the inequality $\int p(x)z^2(x)J(\rho_x)~dx>0$.
Assuming $J(\rho_x)$ is complete positive,  i.e., $J(\rho_x)>0$, let
$z(x)=E'(x) /J(\rho_x)$ we obtain
\begin{equation}
\int p(x)\delta \hat{f}^2~dx \geq \int p(x)\frac{E'^2(x)}{J(\rho_x)}dx
=\int p(x)\frac{[f'(x)+b'(x)]^2}{J(\rho_x)}dx.
\end{equation}
From Eq.~({\ref{eq:msedef}) we then get the lower bound for the mean square error
\begin{equation}
\mathrm{MSE}(\hat{f})\geq \int p(x)\left\{\frac{[f'(x)+b'(x)]^2}
{J(\rho_x)}+b^2(x)\right\}dx. \label{eq:MSE}
\end{equation}
When $b(x)=0$, i.e., for unbiased estimators the bound reduces to a Bayesian
Cram\'{e}r-Rao bound~\cite{Gill95} (another Bayesian QCRB using left logarithmic derivative
is in Ref.~\cite{Yuen1973}) Furthermore, if $f(x)=x$,  the bound reduces
to the well-used Cram\'{e}r-Rao form~\cite{BRAU94}.  If we only consider $f(x)=x$
and take the prior distribution as a uniform one, above bound can be treated as the
quantum version of the biased Cram\'{e}r-Rao bound~\cite{HELS67}. The bound given in Eq.~(\ref{eq:MSE})
vividly  displays the tradeoff between the variance and the bias of the estimate:
at one extreme by letting $b(x)=0$ the unbiased  estimates minimize the term $b^2(x)$,
while the first term is fixed; at the other extreme by letting $b(x)=-f(x)$ we can minimize the
the first term, but now with a fixed bias $b^2(x)=f^2(x)$. The actual minimum of this bound
lies somewhere between these two extremes, which provides a lower bound for all estimators.

To obtain a valid lower bound for all estimators we use the variational principle to find the
optimal $b(x)$ that minimizes the bound in Eq.~(\ref{eq:MSE}), which follows the treatment in Ref.~\cite{Eldar2009}.
Suppose the support of the prior distribution $p(x)$ is in $(a_1, a_2)$, i.e., $p(x)=0$ for any $x$
outside $(a_1,a_2)$. Denote $G(b,x)=p(x)\big\{[f'(x)+b'(x)]^2/J(\rho_x)+b^2(x)\big\},$
and using variation of calculus, the optimal $b(x)$ that minimizes $\int_{a_1}^{a_2} G(b,x)dx$
should satisfy the Euler-Lagrange equation
\begin{equation}
\frac{\partial G}{\partial b}-\frac{\partial}{\partial x}\frac{\partial G}{\partial b'}=0,
\end{equation}
with the Neumann boundary condition $\frac{\partial G}{\partial b'}\big|_{x=a_1}
=\frac{\partial G}{\partial b'}\big|_{x=a_2}=0$.
Substituting the expression of $G(b,x)$ into the equation, one can obtain
\begin{equation}
p(x)b(x)=\frac{\partial}{\partial x}\left[p(x)\frac{f'(x)+b'(x)}{J(\rho_x)}\right], \label{eq:vari0}
\end{equation}
which gives the following differential equation for the optimal $b(x)$
\begin{eqnarray}
& & p(x)[b''(x)+f''(x)]+p'(x)[f'(x)+b'(x)] \nonumber\\
&=& \frac{p(x) J'(\rho_x)}{J(\rho_x)}[b'(x)+f'(x)]+J(\rho_x)p(x)b(x),
\label{eq:vari}
\end{eqnarray}
which can be reorganized and written compactly as
\begin{equation}
J(\rho_x)b(x) = [b'(x)+f'(x)] \frac{\partial}{\partial x}\!\!\left( \ln \frac{ p(x)}{J(\rho_x)}\right)+b''(x)+f''(x),
\label{eq:vari1}
\end{equation}
with boundary conditions $b'(a_1)=-f'(a_1)$ and $b'(a_2)=-f'(a_2)$.
Note that the obtained solution of $b(x)$ may not correspond to an actual bias of an estimator, it
is just used as a tool to get the lower bound~\cite{Young}.
The optimal bias $b(x)$ can then be obtained by solving this equation,
either numerically or analytically.  Next, substituting it back to Eq.~(\ref{eq:MSE}),
one can get a valid lower bound for all estimates.

If the prior distribution $p(x)$ and the quantum Fisher information $J(\rho_x)$ are independent of $x$,
then the equation simplifies to
\begin{equation}
J b(x) =b''(x)+f''(x),  \label{eq:dobb}
\end{equation}
which can be analytically solved. For example consider a uniform prior distribution on $(0,a)$,
and we would like to estimate the unknown parameter itself, i.e., $f(x)=x$.
In this case we can obtain an analytical solution for the optimal bias
\begin{equation}
b(x)=\frac{\cosh\!\left[\sqrt{J}\left(a-x\right)\right]-\cosh(\sqrt{J}x)}{\sqrt{J}\sinh(\sqrt{J}a)}.
\end{equation}
Substituting it back to the right side of the inequality (\ref{eq:MSE}), we obtain a valid lower bound for all estimates
\begin{equation}
\mathrm{MSE}(\hat{x})\geq \frac{1}{J}-\frac{2}{aJ^{3/2}}\tanh\left(\frac{a}{2}\sqrt{J}\right).
\label{eq:MSE1}
\end{equation}
Compare to the quantum Cram\'er-Rao bound, this bound has an extra term which is then always lower.

\section{Examples}

In this section, we give four examples for the valid lower bound.  In the first
three examples, the QFI is independent of the parameter under estimation.
In these examples, taking the prior distribution as uniform, the MSE can be
directly obtained via Eq.~(\ref{eq:MSE1}). However, in some cases, the QFI
is actually dependent on the estimated parameter. The fourth example is such
a case. In this example, the optimal bias has to be solved  via Eq.~(\ref{eq:vari1}).

\begin{figure}[tp]
\centering\includegraphics[width=9cm]{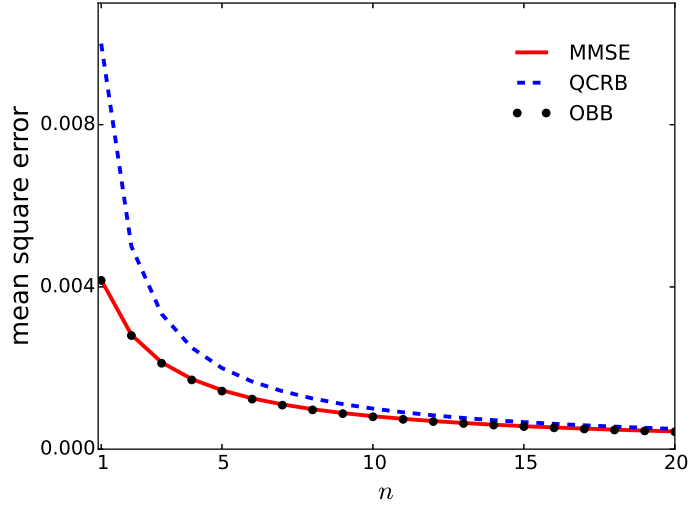}
\caption{(Color online) Mean square error for the minimum mean square error estimator
(MMSE, solid red line, Eq.~(\ref{eq:standard})), optimal biased bound
(OBB, black dots, Eq.~(\ref{eq:mse_eg1}))
and quantum Cram\'er-Rao bound (QCRB, dashed blue line) with different number
of repeated measurements $n$. Here we consider a NOON state of $N=10$ particles.
The prior distribution $p(x)$ is taken as the uniform
distribution on $(0, \pi/10)$.}  \label{fig:N}
\end{figure}

\emph{Example 1}.  As the first example, we consider $N$ spins in the NOON state, $(|00\cdots 0\rangle
+|11\cdots 1\rangle)/\sqrt{2},$ which evolves under the dynamics $U(x)=(e^{-i \sigma_3 xt /2})^{\otimes N}$
(same unitary evolution  $e^{-i \sigma_3 xt/2}$ acts on each of the $N$ spins)
with $\sigma_1=|0\rangle\langle 1|+|1\rangle 0|$, $\sigma_y=-i|0\rangle\langle 1|+i|1\rangle\langle 0|$
and  $\sigma_3 = |0\rangle\langle 0|-|1\rangle\langle 1|$ as Pauli matrices.
After $t$ units of time it evolves to
\begin{equation}
\label{eq:NOON}
|\psi(x)\rangle=\frac{1}{\sqrt{2}}\left(e^{\frac{i}{2}Nxt}|00\cdots 0\rangle
+e^{-\frac{i}{2}Nxt}|11\cdots 1 \rangle\right).
\end{equation}
We can take the time as a unit, i.e.,  $t=1$. This NOON state has the quantum Fisher
information $J=N^2$~\cite{GIOV06}. For $n$ times repeated measurements,
the quantum Fisher information is $nN^2$. If the prior distribution $p(x)$ is uniform on $(0, a)$,
then from Eq.~(\ref{eq:MSE1}), we have
\begin{equation}
\mathrm{MSE}(\hat{x})\geq \frac{1}{n N^2}-\frac{2}{a N^{3} n^{3/2}}\tanh\left(\frac{a}{2} \sqrt{n} N\right).
\label{eq:mse_eg1}
\end{equation}

We will compare these bounds with an actual estimation procedure using the MMSE estimator. 
Consider the measurements in the basis of
$|\psi_0\rangle=(|00\cdots 0\rangle +|11\cdots 1\rangle)/\sqrt{2}$ and
$|\psi_1\rangle=(|00\cdots 0\rangle -|11\cdots 1\rangle)/\sqrt{2},$
which has the measurement results $0$ and $1$
with probability distribution $p_0=|\langle \psi_0|\psi_x
\rangle|^2=\cos^2(Nx/2)$ and $p_1=1-p_0=\sin^2(Nx/2)$.
Assuming the measurement is repeated $n$ times, the probability
that has $k$ outcomes as $1$ is given by
\begin{equation}
p(k|x) = {n \choose k}p^k_1p^{n-k}_0 = {n \choose k}\sin^{2k}\!\left(\frac{Nx}{2}\right)
\!\cos^{2(n-k)}\!\left(\frac{Nx}{2}\right),
\end{equation}
where ${n \choose k}$ is the binomial coefficient. From which we can then obtain the MMSE estimator as explained in the introduction.

To compare the QCRB, MMSE and OBB with the mean square error of this procedure, 
we plot these three quantities as functions of measurement number $n$ in
Fig.~\ref{fig:N}. The solid red, dashed blue lines and black dots in this figure
represent the mean square error for the MMSE estimator, the QCRB and the OBB, respectively.
From which we can see that while QCRB fails to calibrate the performance of the
MMSE estimator, the optimal biased bound provides a valid lower bound.
And from the closeness between the MMSE estimator and the optimal biased
bound, one can gauge that in this case the MMSE estimator is almost optimal.
The bias for the MMSE estimator is also plotted in Fig.~\ref{fig:bias}. It can be seen that when $n$ is small, 
the MMSE estimator is indeed biased, for this reason the QCRB fails to calibrate the performance, while 
when $n$ gets larger, the estimator becomes more unbiased, indicating a transition into the asymptotical 
regime where the QCRB starts to be valid.

\begin{figure}[tp]
\centering\includegraphics[width=9cm]{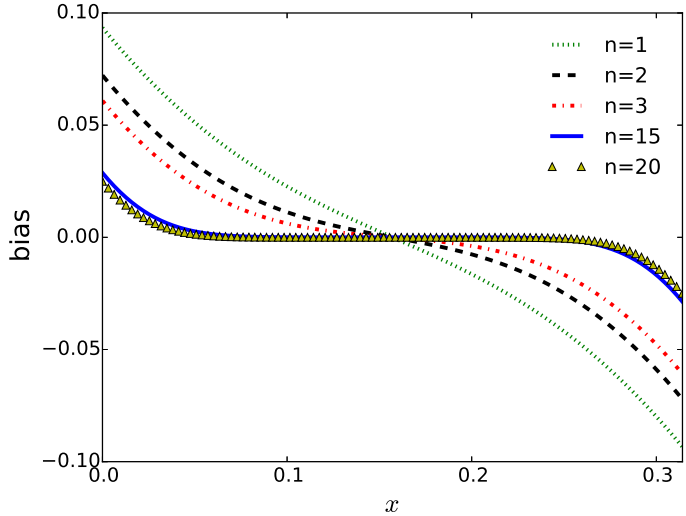}
\caption{(Color online) Bias for posterior mean in minimum mean square error
estimator for different number of measurements. $n=1$: dotted green line; 
$n=2$: dash black line; $n=3$: dash-dotted red line; $n=15$: solid blue line; 
$n=20$: yellow triangulars. Here we consider a NOON state of $N=10$ particles.
The prior distribution $p(x)$ is taken as the uniform distribution on $(0, \pi/10)$.
\label{fig:bias}}
\end{figure}

\emph{Example 2}.  We consider a qubit undergoing an evolution with dephasing noise.
The master equation for the density matrix $\rho$ of the qubit is
\begin{equation}
\dot{\rho}=-i\left[\frac{\sigma_z}{2}x, \rho\right]+\frac{\gamma}{2}
\left(\sigma_z\rho\sigma_z-\rho\right),
\end{equation}
where $\gamma$ is the decay rate and $x$ is the parameter under estimation.
Take the initial state as  $|\psi_0\rangle=(|0\rangle+|1\rangle)/\sqrt{2}$,
then after time $t$, which we normalize to $1$, the evolved state reads
\begin{equation}
\rho_x=\frac{1}{2}\left(\begin{array}{cc}
1 & \eta e^{-ix} \\
 \eta e^{ix} & 1 \\
 \end{array}\right),
 \end{equation}
where $\eta=\exp(-\gamma)$.
The quantum Fisher information in this case is given by $J=\eta^2$.
The quantum Cram\'er-Rao bound for $n$ repeated measurements
then gives
\begin{equation}
\mathrm{MSE}(\hat{x})\geq \frac{1}{n \eta^2}.
\end{equation}

For the optimal biased bound we again takes the prior distribution $p(x)$ as uniform on $(0, \pi)$.
Based on Eq.~(\ref{eq:MSE1}), one can get the optimal biased bound as
\begin{equation}
\mathrm{MSE}(\hat{x}) \geq \frac{1}{n \eta^2}-\frac{2}{\pi(\sqrt{n}\eta)^3}
\tanh\left(\frac{\pi}{2}\sqrt{n}\eta\right). \label{eq:MSE2}
\end{equation}

\begin{figure}[tp]
\centering\includegraphics[width=9cm]{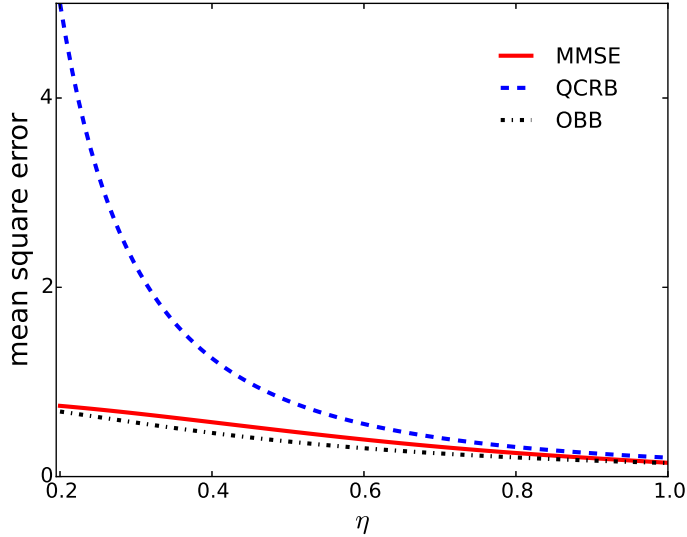}
\caption{(Color online) Mean square error for the minimum mean square error estimator 
(MMSE, solid red line, Eq.~(\ref{eq:standard})), optimal biased bound
(OBB, dash-dotted black line, Eq~(\ref{eq:MSE2})) and quantum Cram\'er-Rao bound
(QCRB, dashed blue line) for a qubit at different rate of dephasing noise $\eta$,
with the measurements number $n=5$.
The prior distribution is taken as the uniform distribution on $(0,\pi)$.}
\label{fig:dephasing}
\end{figure}

We also use this bound to gauge the performance of a measurement scheme, which measures in the basis of
$|\psi_0\rangle=(|0\rangle+|1\rangle)/\sqrt{2}$ and $|\psi_1\rangle=(|0\rangle-|1\rangle)/\sqrt{2}$.
The distributions of the measurement results are given by
\begin{eqnarray}
p(0|x)=\langle\psi_1|\rho_x|\psi_1\rangle=\frac{1+\eta\cos(x)}{2},\\
p(1|x)=\langle\psi_0|\rho_x|\psi_0\rangle=\frac{1-\eta\cos(x)}{2}.
\end{eqnarray}
The probability that has $k$ outcomes as $1$ among $n$ repeated measurements is
$p(k|x)={n \choose k}p^k(1|x)p^{n-k}(0|x).$ Again using the minimum mean square error estimator,
which is given by the posterior mean $\hat{x}(k) = \int p(x|k)x dx $, we can get
the mean square error via Eq.~(\ref{eq:standard}).
In Fig.~\ref{fig:dephasing}, we plotted the mean square error for the MMSE estimator,
the optimal biased bound and quantum Cram\'er-Rao bound at different strength of
dephasing noise.  It can be seen that while the quantum Carm\'er-Rao bound fails to provide
a valid lower bound, the optimal biased bound provides pretty tight bound
at all ranges of dephasing noise, which indicates that the MMSE estimator is close to be
optimal even at the presence of dephasing noises.

\begin{figure}[tp]
\centering\includegraphics[width=9cm]{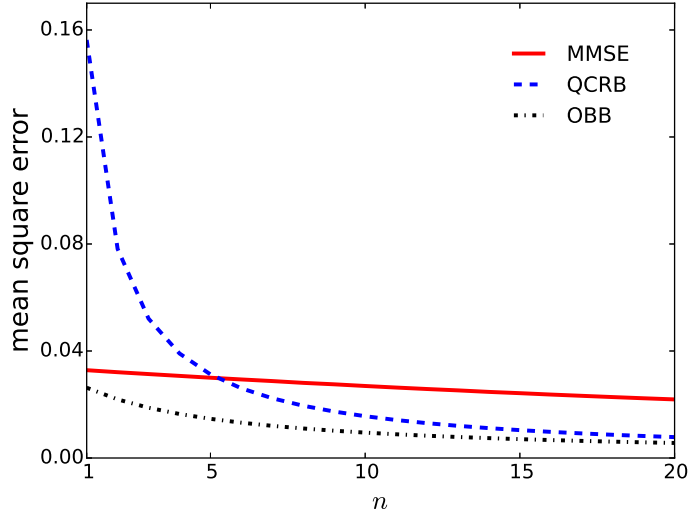}
\caption{(Color online) Optimal biased bound
(OBB, solid red line, Eq.~(\ref{eq:mse_eg3})), quantum Cram\'er-Rao bound
(QCRB, dashed blue line), the minimum mean square error for the MMSE estimator(MMSE, solid red line)
for the phase estimation in the interferometer. Here we consider a SU(2) interferometer with
$n_{\mathrm{A}}=n_{\mathrm{B}}=1$. The prior distribution is uniform in $(0,\pi/5)$.}
\label{fig:inteferometer}
\end{figure}

\emph{Example 3}.  In this example, we consider a SU(2) interferometer described via a unitary
transformation $\exp(-i x S_2)$. Here $S_2$ is a Schwinger operator defined as
$S_{2}=\frac{1}{2i}(a^{\dagger}b-b^{\dagger}a)$ with $a(a^{\dagger})$, $b(b^{\dagger})$
the annihilation (creation) operators for ports A and B. $x$ is the parameter under
estimation.  Now we take the import state as  a coherent state $|\beta\rangle$ for port A and  a cat state
$\mathcal{N}_{\alpha}(|\alpha\rangle+|-\alpha\rangle)$ for port B.  Here
$\mathcal{N}^2_{\alpha}=1/(2+2e^{-2|\alpha|^2})$ is the normalization number.
Taking into account the phase-matching condition, the quantum Fisher information
for $x$ in this case is in the form~\cite{JLiu2013}
\begin{equation}
J = 2 n_{\mathrm{A}}n_{\mathrm{B}}+n_{\mathrm{A}}
+n_{\mathrm{B}}+2n_{\mathrm{A}}|\alpha|^2,
\end{equation}
where $n_{\mathrm{A}}=|\beta|^2$ and $n_{\mathrm{B}}=
|\alpha|^2 \tanh|\alpha|^2$
are photon numbers in port A and B. Based on above expression, the quantum Fisher information $J$ is independent of $x$.
Thus, for the optimal biased estimation, the mean square error $\mathrm{MSE}(\hat{x})$
satisfies Eq.~(\ref{eq:MSE1}). The maximum Fisher information with respect to $n_{\mathrm{A}}$ and
$n_{\mathrm{B}}$ for a fixed yet large total photon number
in this case  can be achieved when photon numbers for both ports are equal,
which is $J_{\mathrm{m}}=N^2+N$~\cite{JLiu2013}, with $N$ the total photon number
in the interferometer.  Using the optimal biased bound and taking the prior distribution
as uniform on $(0, a)$, for $n$ times repeated  measurements, $\mathrm{MSE}(\hat{x})$ then satisfies
\begin{equation}
\mathrm{MSE}(\hat{x}) \geq  \frac{1}{n J}-\frac{2}{a(n J)^{3/2}}
\tanh\left(\frac{a}{2}\sqrt{n J}\right). \label{eq:mse_eg3}
\end{equation}

Figure~\ref{fig:inteferometer} shows the quantum Cram\'{e}r-Rao bound
(dashed blue line), the optimal biased bound (dash-dotted black line) and the minimum mean square
error for the MMSE estimator (solid red line). The prior distribution taken as uniform in $(0,\pi/5)$. In this figure,
$n_{\mathrm{A}}=n_{\mathrm{B}}=1$. For the MMSE estimator, we measure along the state
$|11\rangle$. We can see that the optimal biased bound provides a valid lower bound at all range of $n$, however 
the gap between the mean square error of the MMSE estimator and the bound indicates that the measurement along 
the state $|11\rangle$ may not be optimal.

\emph{Example 4}. The quantum Fisher information in above examples is independent of
the estimating parameter $x$. We give another example with the quantum
Fisher information depending on $x$.

Consider a qubit system with the Hamiltonian
\begin{equation}
H = \frac{B}{2}\left(\sigma_1 \cos x+\sigma_3 \sin x\right),
\end{equation}
which describes the dynamics of a qubit under a magnetic field in the $XZ$ plane, the interested parameter denotes the direction of the magnetic field.
The quantum Fisher information of this system has been recently studied with various methods~\cite{Yuan2015, JLiu2015, Pang2014}.
For a pure initial state $(|0\rangle+|1\rangle)/\sqrt{2}$, the quantum Fisher information is given by (with the evolution time normalized as $t=1$)
\begin{equation}
J(x) = 4\sin^2\left(\frac{B}{2} \right)\left[1-\cos^2 \left(\frac{B}{2}\right)\sin^2 x \right],
\end{equation}
which depends on $x$. In this case, we have to solve Eq.~(\ref{eq:vari1}).  Like previous examples,
we take the prior distribution $p(x)$ as uniform on $(0, \pi/2)$. If we take $B=\pi/2$,
with $n$ repeated measurements, $J=n(2-\sin^2 x)$, then
Eq.~(\ref{eq:vari1}) reduces to
\begin{equation}
n(2-\sin^2 x)b''+\sin(2 x)b'=(2-\sin^2 x)^2 b  -\sin(2 x).
\end{equation}
This equation can be numerically solved and by substituting the obtained $b(x)$ into Eq.~(\ref{eq:MSE}),
the optimal biased bounds can be obtained which is plotted in Fig.~\ref{fig:J_x}.

\begin{figure}[tp]
\centering\includegraphics[width=9cm]{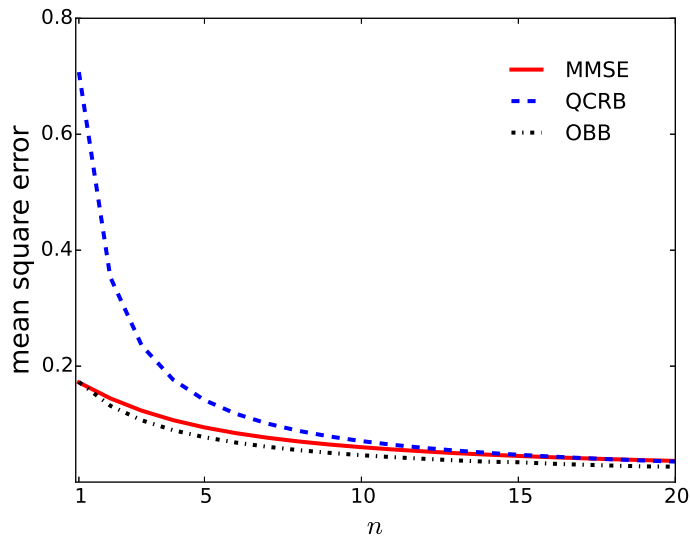}
\caption{(Color online) Mean square error for minimum mean square error estimator 
(MMSE, solid red line, Eq.~(\ref{eq:standard})), the optimal biased bound
(OBB, dash-dotted black line) and quantum Cram\'er-Rao bound (QCRB, dashed blue line) as a function
of measurement number $n$.  Here we consider a qubit under a magnetic field in the XZ plane.
The prior distribution is taken as uniform in $(0,\pi/2)$.}
\label{fig:J_x}
\end{figure}

Again we use this bound to gauge the performance of a
measurement scheme which takes measurements along
$|\psi_0\rangle=(|0\rangle+|1\rangle)/\sqrt{2}$ and
$|\psi_1\rangle=(|0\rangle-|1\rangle)/\sqrt{2}$.
The probability distribution of the measurement results are given by
\begin{equation}
p(1|x) = \sin^2\left(\frac{B}{2}\right)\sin^2 x,
\end{equation}
and $p(0|x)=1-p(1|x)$.
When $B$ equals to $\pi/2$, above probability reduces to $p(1|x)=(\sin^2 x)/2$.
The probability having $k$ outcomes as $1$ among $n$ repeated measurements is
$p(k|x)={n \choose k}p^k(1|x)p^{n-k}(0|x).$ Using the posterior mean as the estimator,
we can obtain the mean square error for the MMSE estimator which is also plotted in Fig.~\ref{fig:J_x}.
From this figure, one can again see that while the quantum Cram\'er-Rao bound (dashed blue line) 
fails to gauge the performance of the MMSE estimator (solid red line), the optimal biased bound (dash-dotted black line) 
provides a valid lower bound and from the closeness between the mean square error of the MMSE estimator and
the optimal biased bound, one can tell that the MMSE estimator is a good estimator here.

\section{Summary}
The optimal biased bound provides a valid
lower bound for all estimators, either biased or unbiased. It can thus be used to
calibrate the performance of all estimators in quantum parameter estimation. Asymptotically the widely
used quantum Cram\'er-Rao bound provides a lower bound for quantum
parameter estimation, however in practice the number of measurements
are often constrained by resources, and it is hard to tell when quantum
Cram\'er-Rao bound applies. From the difference between the optimal
biased bound and quantum Cram\'er-Rao bound it also provides a way to
estimate the number of measurements needed to enter the asymptotical regime.

\end{document}